\documentstyle[epsfig,sprocl]{article}

\begin{document}
\title{$pp$ Elastic Scattering at LHC and Signature of Chiral Phase Transition at Large $|t|$\footnote{Presented at the Fifth Workshop on QCD, Villefranche, France, Jan. 2000 (to be published in Proceedings of the Workshop)}}
\author{M.M. Islam}
\address{Department of Physics\\
University of Connecticut\\
Storrs, CT 06269 \\
U.S.A.}
\maketitle
\thispagestyle{empty}
\begin{abstract}
A model of $pp$ and $\bar{p}p$ elastic scattering developed previously to analyze ISR and SPS Collider data is extended to predict $pp$ elastic differential cross section at LHC at c.m. energy of $14$ TeV and momentum transfer range $|t|=0$ - $10$ GeV$^2$.  Role of the gauged linear $\sigma$-model as an underlying field theory model describing nucleon structure and elastic scattering is discussed.  Possibility of finding evidence of a chiral phase transition at large $|t|$ in the proposed TOTEM project at LHC is pointed out.
\end{abstract}
\section{Introduction}
High energy elastic $pp$ and $\bar{p}p$ scattering at the ISR and SPS Collider in the c.m. energy range $\sqrt{s} = 23$ - $630$ GeV have been analyzed by us in a model where the nucleon has an inner core and an outer cloud~\cite{islam1}.  Recently, we have been able to extend this analysis to the asymptotic energy region~\cite{islam2}.  This has allowed us to predict $pp$ elastic differential cross section at the LHC energy of $\sqrt{s} = 14$ TeV and momentum transfer range $|t| = 0$ - $10$ GeV$^2$.

Previous investigation has shown that in the gauged linear $\sigma$-model the nucleon core can be identified as a soliton arising from the topological baryonic charge distribution, and the outer cloud can be understood as a condensed quark-antiquark ground state analogous to a superconducting ground state~\cite{islam3}.

The gauged linear $\sigma$-model furthermore indicates that at some small distance, equivalently, at some large $|t|$ --- a chiral phase transition from the nonperturbative region to the perturbative region will occur.  As we will see, this leads to a change in the behavior of $\frac{d\sigma}{dt}$ from an exponential fall-off to a power fall-off, and therefore to a distinct change in the slope of $\frac{d\sigma}{dt}$.  Currently, a project to measure $pp$ elastic differential cross section at LHC is being studied.  This project, called TOTEM (acronym for Total and Elastic Measurement)~\cite{totem} will provide detailed features of $pp$ elastic $\frac{d\sigma}{dt}$ from $|t| = 0$ - $10$ GeV$^2$. Furthermore, the large design luminosity of LHC will allow in this project measurement of $\frac{d\sigma}{dt}$ in the range $|t| = 10$ - $15$ GeV$^2$.  The TOTEM project, therefore, will be able to observe any distinct change in the slope of $\frac{d\sigma}{dt}$ at large $|t|$, and thereby will be able to test whether a chiral phase transition does, in fact, occur.

\section{$pp$ Elastic Scattering at LHC}

We view high energy elastic scattering as primarily due to two processes:  one is a glancing collision where the outer cloud of one nucleon interacts with that of the other and gives rise to diffraction scattering; the other is a hard collision where one nucleon core scatters off the other core via vector meson $\omega$-exchange, while their outer clouds overlap and interact independently.  Diffraction dominates the small $|t|$ region, but as the momentum transfer increases --- hard scattering takes over.

We describe the diffraction amplitude using an impact parameter representation:
\begin{equation}
T_D(s,t) = ipW \int_0^{\infty} bd b \; J_0(bq) \Gamma_D^{+}(s,b)
\label{eq:impact}
\end{equation}
with the profile function
\begin{equation}
\Gamma_D^{+}(s,b) = g(s) \left[ \frac{1}{1+e^{\frac{b - R(s)}{a(s)}}} + \frac{1}{1+e^{- \frac{b + R(s)}{a(s)}}} - 1 \right];
\label{eq:profile}
\end{equation}
here $q = \sqrt{|t|}$, $R(s) = R_0 + R_1(\ln s - i\pi/2)$, $a(s) = a_0 + a_1(\ln s - i\pi/2)$.  $g(s)$ is a complex energy dependent parameter that satisfies the crossing symmetry requirement: $g^{*}(se^{i\pi}) = g(s)$.  Our hard scattering amplitude has the form
\begin{equation}
T_{H}(s,t) \sim e^{i \chi_D^+(s,0)} s \frac{F^2(t)}{m_{\omega}^2 - t} \mbox{ .}
\label{hard-scatter}
\end{equation}
It shows that $\omega$ behaves as an elementary vector meson in our model.  The factor of $s$ in Eq. (\ref{hard-scatter}) originates from spin $1$ of $\omega$; the t-dependence, which is the product of two form factors and the $\omega$ propagator, indicates that $\omega$ acts like an elementary meson probing two density distributions.  Such a behavior of $\omega$ cannot be understood in the usual Regge pole model.  However, it can be easily understood in the gauged $\sigma$-model, where $\omega$ couples to the topological baryonic current as a gauge boson (just the way photon couples to the electromagnetic current).  The factor $e^{i \chi_D^+(s,0)}$ in (\ref{hard-scatter}) represents an absorptive correction due to diffraction.

\begin{figure}[b]
\begin{center}
\epsfig{file=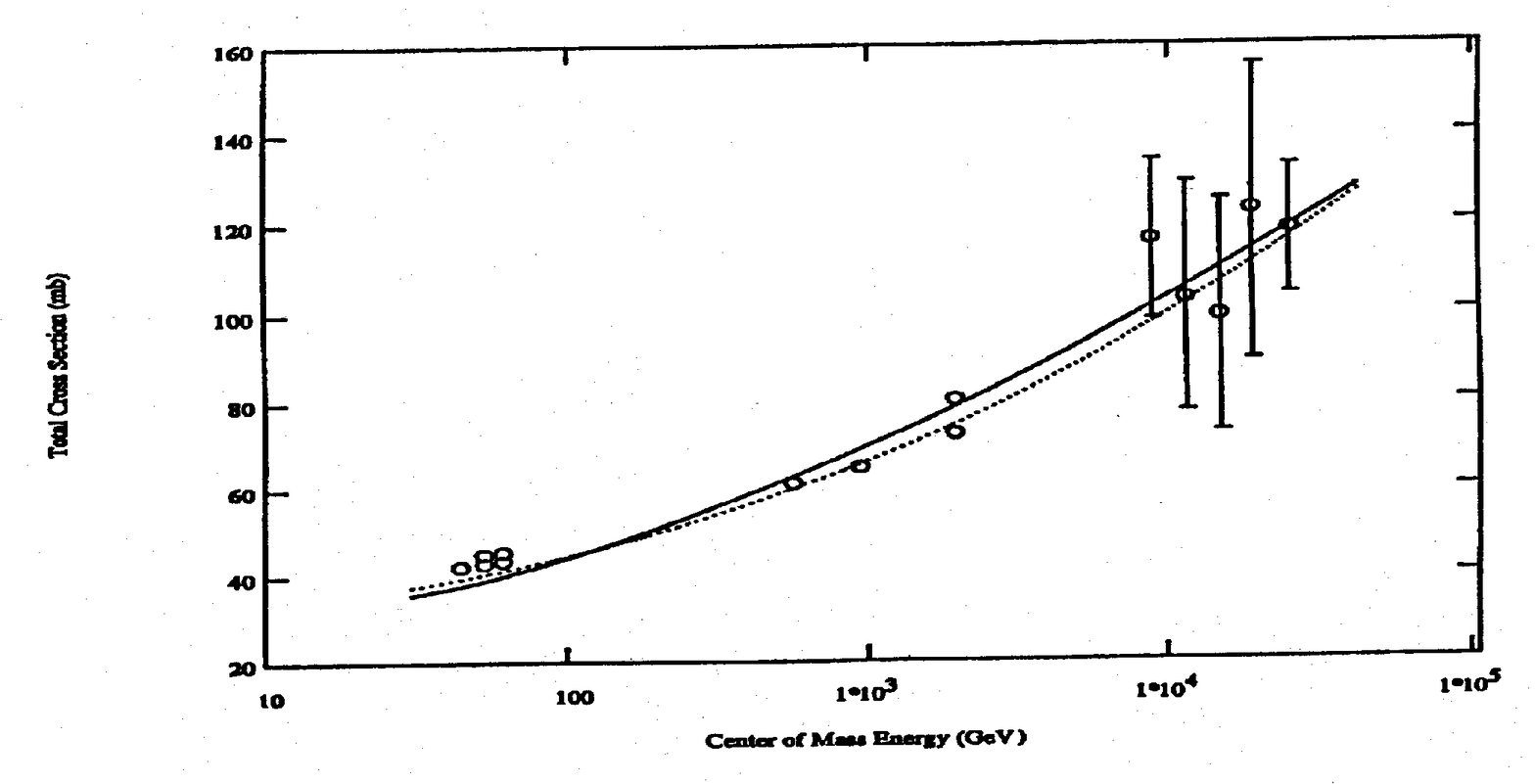,width=4in}
\end{center}
\caption{Solid line represents our calculated total cross section as a function of energy using the parameters given in Ref. 2.  Dotted line represents asymptotic behavior of total cross section given by Augier et. al.}
\label{fig:one}
\end{figure}

\begin{figure}[b]
\begin{center}
\epsfig{file=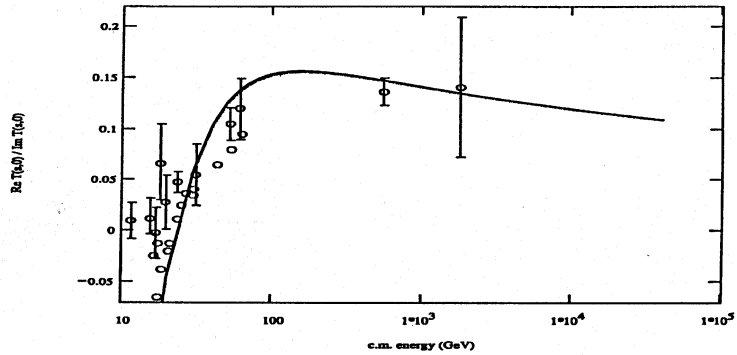,width=4in}
\end{center}

\caption{Solid line represents our calculated $\rho(s) = \mbox{Re} T(s,0)/\mbox{Im} T(s,0)$ as a function of energy using the parameters given in Ref. 2.  As in the case of total cross section, $\rho(s)$ for $pp$ and $\bar{p}p$ essentially overlap.}
\label{fig:two}
\end{figure}

\begin{figure}[b]
\begin{center}
\epsfig{file=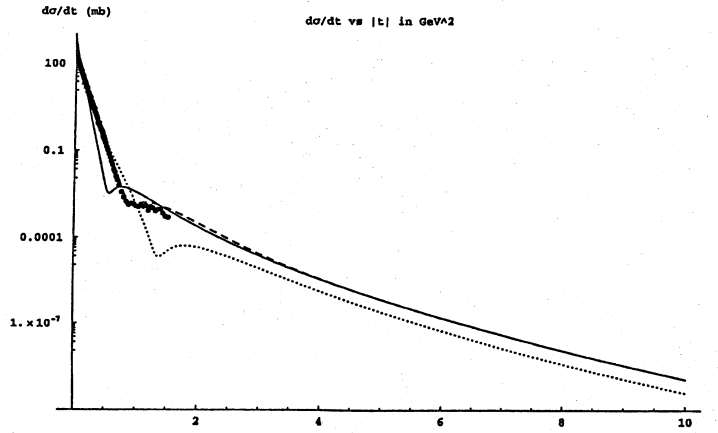,width=4in}
\end{center}
\caption{Solid line represents our predicted differential cross section for $pp$ elastic scattering at the Large Hadron Collider at $\sqrt{s} = 14$ TeV.  Dashed line is our fit of the measured $\bar{p}p$ differential cross section at $\sqrt{s} = 546$ GeV using the parameters given in Ref. 2.  Experimental data at $546$ GeV (solid circles) are from Ref. 6.  Also plotted for comparison is the $pp$ differential cross section at $\sqrt{s} = 53$ GeV using the parameters of Ref. 1 (dotted line).}
\label{fig:three}
\end{figure}

To extend previous calculations to the LHC energy, we needed a way to determine the $s$ dependence of $g(s)$.  Accurate model independent analysis by Kundrat and Lokajicek~\cite{kundrat} shows that absorption due to diffraction at $b=0$, i.e., $\left| e^{i \chi_D^+(s,0)} \right|$ is finite at high energy and decreases slowly with increasing $s$.  This led us~\cite{islam2} to consider the following parameterization for $e^{i \chi_D^+(s,0)}$:
\begin{equation}
e^{i \chi_D^+(s,0)} = \eta_0 + \frac{1}{c_0 + c_1 (\ln s - i\pi/2)} \mbox{ .}
\label{eq:parametrize}
\end{equation}
Since $g(s)$ is related with $e^{i\chi_D^+(s,0)} = 1 - \Gamma_D^+(s,0)$, Eq. (\ref{eq:parametrize}) expresses the energy dependence of $g(s)$ in terms of energy independent parameters $\eta_0$, $c_0$, $c_1$.  We also took into account the energy dependence of $\Gamma_D^-(s,0)$ in a similar fashion.  The net result was that all our parameters were now energy independent.  We determined them by fitting the asymptotic behavior of total cross section $\sigma_{\textstyle{tot}}(s)$ and of $\rho(s) = \mbox{Re}T(s,0)/\mbox{Im}T(s,0)$ (Figs. \ref{fig:one} and \ref{fig:two}).  Also, we required a good fit of $\bar{p}p$ elastic $\frac{d\sigma}{dt}$ at $\sqrt{s} = 546$ GeV, which was measured at the SPS Collider~\cite{bozzo}.  Our predicted $pp$ elastic differential cross section at the LHC energy $\sqrt{s} = 14$ TeV is shown in Fig. \ref{fig:three} (solid line).  The dashed line is our fit of $\bar{p}p$ $\frac{d\sigma}{dt}$ at $\sqrt{s} = 546$ GeV.  The dotted line shows previous calculation by us~\cite{islam1} of $pp$ $\frac{d\sigma}{dt}$ at c.m. energy $53$ GeV and is given for comparison~\cite{comment}.

Examining Fig. \ref{fig:three}, we notice several interesting features.  First, the differential cross section at $|t|=0$ increases, the diffraction peak shrinks, and the interference dip moves closer to the forward direction and rises up.  Second, for $|t| > 1$ GeV$^2$, the differential cross section behaves as $\frac{d\sigma}{dt} \sim e^{-a\sqrt{|t|}}$ (known as Orear fall-off) and falls smoothly several decades.  This is in contrast to the impact factor model of Bourrely et al.~\cite{bourrely} and Regge pole plus cut model of Desgrolard et al.~\cite{desgrolard}, which predict prominent oscillations in the differential cross section~\cite{buenerd}.  Third, for $|t| > 1$ GeV$^2$, $\frac{d\sigma}{dt}$ at c.m. energies $546$ GeV and $14$ TeV coincide showing that $\frac{d\sigma}{dt}$ has little energy dependence, only $|t|$ dependence.  Such a behavior was predicted by Donnachie and Landshoff~\cite{donnachie} using a triple gluon exchange model.  However, they predict $\frac{d\sigma}{dt} \sim \frac{1}{t^8}$, which is very different from what we predict.  Finally, we notice that $\frac{d\sigma}{dt}$ seems to continue its exponential fall-off all the way to $|t|=10$ GeV$^2$ and beyond.  In reality, the underlying gauged linear $\sigma$-model indicates that at some large $|t|$ value, say, $|t| \simeq 8$ GeV$^2$ the behavior of $\frac{d\sigma}{dt}$ will change to a power fall-off: $\frac{d\sigma}{dt} \sim \frac{1}{t^{10}}$, and this is connected with a chiral phase transition.  Let us now examine how the gauged linear $\sigma$-model leads to such a prediction.

\section{Gauged Linear $\sigma$-Model and Signature of Chiral Phase Transition at Large $|t|$}

To introduce the gauged linear $\sigma$-model, we start with the Gell-Mann-Levy $\sigma$-model that has $SU(2)_L \times SU(2)_R$ global symmetry and spontaneous breaking of chiral symmetry:
\begin{eqnarray}
{\cal L} & = & \bar{\psi} i \gamma^\mu \partial_\mu \psi + 
  \frac{1}{2} \left[ \partial_\mu \sigma \partial^\mu \sigma + 
    \partial_\mu \vec{\pi} \cdot \partial^\mu \vec{\pi} \right] \nonumber \\
  & & - g \bar{\psi} \left[ \sigma + i \vec{\tau} \cdot \vec{\pi} \gamma^5 \right] \psi
   - \lambda \left( \sigma^2 + \vec{\pi}^2 - f^2_{\pi} \right)^2.
\label{eq:lagrange}
\end{eqnarray}
Next we introduce a scalar-isoscalar field $\zeta(x)$ and a unitary field $U(x)$ in the following way
\begin{equation}
\sigma(x) + i\vec{\tau}(x) \cdot \vec{\pi}(x) = \zeta(x) U(x).
\end{equation}
$\zeta(x)$ is the magnitude of the fields $\sigma(x)$ and $\vec{\pi}(x)$: $\zeta^2(x) = \sigma^2(x) + \vec{\pi}^2(x).$  $U(x)$ is given by: $U(x) = e^{i \vec{\tau} \cdot \vec{\phi}(x)/f_{\pi}}$, where $\vec{\phi}(x)$ is the Goldstone pion field and $f_{\pi}$ is the pion decay constant ($f_{\pi} \simeq 93$ MeV).  Introducing right and left fermion fields $\psi_{R,L} = \frac{1}{2}(1 \pm \gamma^5) \psi$, the Lagrangian (\ref{eq:lagrange}) can be written as 
\begin{eqnarray}
{\cal L} & = & \bar{\psi}_R i \gamma^\mu \partial_\mu \psi_R +  \bar{\psi}_L i \gamma^\mu \partial_\mu \psi_L + \frac{1}{2} \partial_\mu \zeta \partial^\mu \zeta \nonumber \\
& & + \frac{1}{4} \zeta^2 \textstyle{tr} \left[ \partial_\mu U \partial^\mu U^\dagger \right] - g \zeta \left( \bar{\psi}_L U \psi_R + \bar{\psi}_R U^\dagger \psi_L \right) - \lambda \left( \zeta^2 - f^2_{\pi} \right)^2.
\label{eq:newlagrange}
\end{eqnarray}
Under right and left global transformations, $\psi_R(x) \rightarrow R\psi_R(x)$, $\psi_L(x) \rightarrow L\psi_L(x)$, $U(x) \rightarrow LU(x)R^\dagger$, and the Lagrangian (\ref{eq:newlagrange}) remains invariant.  When the scalar field $\zeta(x)$ is replaced by its vacuum value $f_{\pi}$, the Lagrangian (\ref{eq:newlagrange}) becomes a nonlinear $\sigma$-model.

The model (\ref{eq:newlagrange}) can be gauged by a massive Yang-Mills approach, or by the hidden local gauge approach of Bando et al.  We adopt the latter~\cite{islam12} and write $U(x) = \xi_L^\dagger(x) \xi_R(x)$, where $ \xi_L(x)$ and $\xi_R(x)$ are $SU(2)$ valued fields which transform in the following way under $\left[ SU(2)_L \times SU(2)_R \right]_{\textstyle{global}} \times$ \\
 $\left[ SU(2) \times U(1) \right]_{\textstyle{local(hidden)}}$ symmetry:
\begin{equation}
\xi_R(x) \rightarrow h(x) \xi_R(x)R^\dagger \mbox{,  } \xi_L(x) \rightarrow h(x) \xi_L(x) L^\dagger;
\label{eq:xi}
\end{equation}
$h(x)$ is an element of $\left[ SU(2) \times U(1) \right]_{\textstyle{local}}$.  Eq. (\ref{eq:xi}) shows that $U(x) = \xi_L^\dagger(x) \xi_R(x) \rightarrow L U(x) R^\dagger$, as required, so that the global symmetry is maintained.  The hidden local symmetry is next gauged by introducing vector mesons $\rho$ and $\omega$ as gauge bosons.  Furthermore, in the fermion-scalar sector, new fermionic variables $\psi_R^\circ(x) = \xi_R(x) \psi_R(x)$ and $\psi_L^\circ(x) = \xi_L(x) \psi_L(x)$ are introduced.  These variables are invariant under global transformations and transform only under hidden local symmetry.  The path integral representation of the partition function of the model now takes the form~\cite{islam12}:
\begin{equation}
\textstyle{Tr}\left[ e^{-iHt} \right] = \frac{1}{{\cal N}} \int \; {\cal D}\zeta \; {\cal D}\xi \; {\cal D}\psi \; {\cal D}\bar{\psi} \; e^{i\int d^4x \left[ {\cal L}_\pi(\xi,\zeta,V) + {\cal L}_{q,\zeta}(\psi_0,\bar{\psi}_0,\zeta,V) \right]}
\label{eq:model} 
\end{equation}
in the unitary gauge $(\xi_L^\dagger(x) = \xi_R(x)=\xi(x)$, $U(x)=\xi^2(x))$.  The subscript $q$ of ${\cal L}_{q,\zeta}$ is to indicate that we consider the fermions to be quarks (of course, dressed quarks of an effective theory and not the current quarks of perturbative QCD).  Next we consider changing the fermion measure in (\ref{eq:model}) from ${\cal D}\psi {\cal D}\bar{\psi}$ to ${\cal D}\psi^\circ {\cal D}\bar{\psi}^\circ$.  This gives rise to a Jacobian that can be written as $e^{i\Gamma[\xi,V]}$, where $\Gamma[\xi,V] = \Gamma_{{WZW}}[\xi,V]$ is the gauged Wess-Zumino-Witten action.  In the simplest approximation, the Lagrangian associated with $\Gamma_{WZW}$ is ${\cal L}_{WZW} = -(N_cg_V/2)\omega_\mu B^\mu$, where $N_c=3$ is the number of colors,  $g_V$ is the vector meson coupling constant, and $B^\mu$ is the topological baryon number current $B^\mu = \frac{1}{24\pi^2} \varepsilon^{\mu \nu \rho \sigma} \textstyle{tr} \left[ U^\dagger \partial_\nu U U^\dagger \partial_\rho U U^\dagger \partial_{\sigma} U \right]$.  The partition function (\ref{eq:model}) now becomes
\begin{eqnarray}
\textstyle{Tr}\left[ e^{-iHt} \right] & = & \frac{1}{{\cal N}} \int \; {\cal D}\zeta \; {\cal D}\xi \; e^{i \int d^4x \left[ {\cal L}_\pi(\xi,\zeta,V) + {\cal L}_{WZW}\right]}  \nonumber \\
& & \times \int \; {\cal D}\psi^\circ \; {\cal D}\bar{\psi}^\circ \; e^{i\int d^4x {\cal L}_{q,\zeta}(\psi^\circ,\bar{\psi}^\circ,\zeta,V)} \mbox{ .}
\label{eq:model2} 
\end{eqnarray}
We now consider the approximation that the scalar field $\zeta(x)$ may be replaced by its vacuum value $f_{\pi}$ in the pion sector.  This leads to 
\begin{eqnarray}
\textstyle{Tr}\left[ e^{-iHt} \right] & \simeq & \frac{1}{{\cal N}} \int \; {\cal D}\xi \; e^{i \int d^4x \left[ {\cal L}_\pi(\xi,f_{\pi},V) + {\cal L}_{WZW}\right]} \nonumber \\
& &  \times \int \; {\cal D}\zeta \; {\cal D}\psi^\circ \; {\cal D}\bar{\psi}^\circ \; e^{i\int d^4x {\cal L}_{q,\zeta}(\psi^\circ,\bar{\psi}^\circ,\zeta,V)}
\label{eq:newmodel} \mbox{ .}
\end{eqnarray}
The important point to note is that the partition function has become the product of two partition functions: one for the pion sector and the other for the quark-scalar sector.  The pion sector with the WZW action included represents the gauged nonlinear $\sigma$-model that describes the nucleon as a topological soliton, i.e., as a Skyrmion. This model has been extensively studied for more than a decade by many groups and has been very successful in describing low energy properties of the nucleon~\cite{zhang,schechter}.  However, two assumptions are implicit in the usual nonlinear $\sigma$-model.  First, the scalar field can be replaced by its vacuum value $f_{\pi}$ from the very beginning, so that there is no scalar degree of freedom.  Second, all the important interactions are in the pion sector; the only important interaction that comes from the fermion or quark sector is the WZW action.  Consequently, this model depicts the nucleon as a topological soliton lying in a noninteracting Dirac sea (Fig. 4a).  On the other hand, we may regard approximating $\zeta$ by its vacuum value $f_{\pi}$ in the pion sector as reasonable, but not neglecting its important interaction in the quark-scalar sector, because the latter interaction makes the quarks massive and leads to the spontaneous breaking of chiral symmetry.  The physical picture that emerges from Eq. (\ref{eq:newmodel}) then is that the topological soliton lies in an interacting Dirac sea, where left and right quarks interact via the scalar field (Fig. 4b).  For a scalar field with a critical behavior, and by this I mean it is zero at small distances, but rises sharply at some distance $r=R$ to its vacuum value $f_{\pi}$ (Fig. 4c), the combined energy of the interacting Dirac sea and the scalar field is considerably less than that of the noninteracting Dirac sea~\cite{islam3}.  The system therefore makes a transition to an interacting ground state and reduces its total energy significantly.  The phenomenon is analogous to superconductivity.  The emergence of a $q\bar{q}$ condensed ground state surrounding the topological soliton shows that it can be identified as the outer cloud of the nucleon.  The gauged linear $\sigma$-model thus serves as an underlying field theory model of our phenomenological description of nucleon structure and elastic scattering. However, the behavior of the scalar field shown in Fig. 4c also leads to an important consequence.

In high energy elastic scattering at a momentum transfer $Q = \sqrt{|t|}$, a nucleon probes the other nucleon at an impact parameter or transverse distance $b \simeq 1/Q$.  Fig. 4c shows that as long as $1/Q > R$, one nucleon is probing the other in the region where $\zeta=f_{\pi}$, i.e., in the nonperturbative region where we have the soliton description and the $q\bar{q}$ condensed ground state.  However, if $Q$ becomes so large that $1/Q < R$, then one nucleon probes the other in the region where the scalar field vanishes.  The quarks are then massless, and we are in a perturbative regime.

High energy elastic $pp$ scattering in the perturbative regime at large $Q^2$ has been investigated by Sotiropoulos and Sterman (SS)~\cite{sotiropoulos}.  Their result is 
\begin{equation}
\left( \frac{d\sigma}{dt} \right)_{pp} \simeq \left( \frac{d\sigma}{dt} \right)_{qq} \frac{1}{t^8} \mbox{ ,}
\label{eq:diff-cross-section}
\end{equation}
where the quark-quark scattering is due to one valence quark scattering off another valence quark; the $t^{-8}$ factor arises from the wave funtion requirement that spectator valence quarks have to be within a distance $1/Q$ of the colliding valence quark.  SS concluded that the net result is
\begin{displaymath}
\left( \frac{d\sigma}{dt} \right)_{pp} \sim \frac{1}{t^{10}},
\end{displaymath}
same as given by quark counting rules.  From our point of view, this means --- as the momentum transfer $Q$ increases, there will be a critical value $Q_0 \simeq 1/R$ beyond which $\frac{d\sigma}{dt}$ will rapidly tend to a power fall-off.  Thus, a distinct change in the slope of $\frac{d\sigma}{dt}$ will occur, which will represent a transition from the nonperturbative regime to a perturbative regime and will be a signature of the chiral phase transition.

Experimentally, a change in $\frac{d\sigma}{dt}$ slope was observed at ISR~\cite{dekerret} at c.m. energy $53$ GeV and $|t| \raisebox{-.6ex}{ $\stackrel{>}{\sim}$ } 8$ GeV$^2$.  This would mean $Q_0^2 \simeq 8$ GeV$^2$ and $R \simeq 0.07$ F.  However, the change in slope could not be established definitively, because of low statistics. The TOTEM project, on the other hand, will be able to measure accurately $pp$ elastic differential cross section up to momentum transfer as large as $|t| \simeq 15$ GeV$^2$, because of the high luminosity of LHC.  Therefore, this experiment has the potential to verify whether a chiral phase transition occurs at small distance.

\section{Concluding Remarks}

We conclude with the following remarks:

\begin{enumerate}
\item Our previous analyses of high energy $pp$ and $\bar{p}p$ elastic scattering can now be extended in the asymptotic region.  This is accomplished by requiring that the paramters of the model describe quantitatively the asymptotic behavior of total cross section and of $\rho$, and fit the measured $\bar{p}p$ differential cross section at $\sqrt{s} = 546$ GeV.  Determination of the parameters allows us to predict $pp$ differential cross section at LHC at the c.m. energy $14$ TeV and momentum transfer range $|t| = 0$ - $10$ GeV$^2$.
\item The gauged linear $\sigma$-model provides an effective field theory framework of our phenomenological description of the nucleon structure and elastic scattering.  It identifies the nucleon core as a topological soliton and the outer cloud as a quark-antiquark condensed ground state.  Furthermore, it predicts a phase transition at some small distance ($R \simeq 0.07$ F), equivalently, at some large momentum transfer ($Q^2 \raisebox{-.6ex}{ $\stackrel{>}{\sim}$ } 8$ GeV$^2$).
\item The proposed TOTEM project, by measuring accurately $pp$ elastic differential cross section at LHC in the range $|t| = 0$ - $10$ GeV$^2$, will be able to verify the features associated with diffraction scattering and the soliton structure, as shown in Fig. 3.  Furthermore, by extending $\frac{d\sigma}{dt}$ measurement to $|t| \simeq 15$ GeV$^2$, it will be able to establish any distinct change in the slope of $\frac{d\sigma}{dt}$ for large $|t|$ and thereby determine whether a chiral phase transition does, in fact, occur.
\end{enumerate}

\begin{figure}[b]
\begin{center}
\epsfig{file=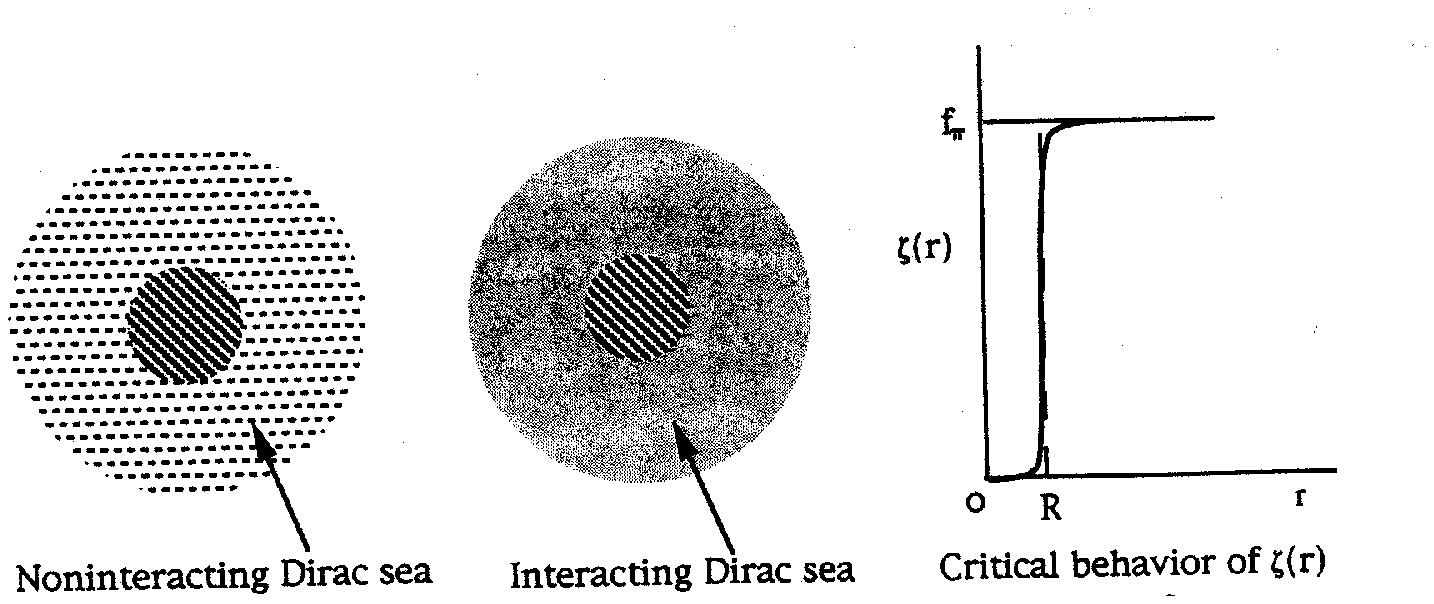,width=4.5in}
\mbox{Figure 4a  \hspace{.7in} Figure 4b \hspace{.7in} Figure 4c}\\
\end{center}
\end{figure}

\end{document}